\documentclass[aps,shownopacs,superscriptaddress,nofootinbib,preprint,11pt]{revtex4}
\usepackage{hyperref}
\usepackage{amsmath}
\usepackage{float}
\usepackage{graphics}
\usepackage{graphicx}
\usepackage{epsfig}
\usepackage{psfrag}
\usepackage{color}
\usepackage{slashed}
\usepackage{amsfonts}
\usepackage{amssymb}

\newcommand*{\be}{\begin{equation}}
\newcommand*{\ee}{\end{equation}}
\newcommand*{\bea}{\begin{eqnarray}}
\newcommand*{\eea}{\end{eqnarray}}

\newcommand{\comment}[1]{}

\newcommand{\cref}[1]{Chapter~\ref{c.#1}}

\newcommand{\barr}{\begin{eqnarray}}
\newcommand{\earr}{\end{eqnarray}}
\def\nn{\nonumber \\}

\def\beq{\begin{equation}}
\def\eeq{\end{equation}}
\def\bea{\begin{eqnarray}}
\def\eea{\end{eqnarray}}
\def\ba{\begin{array}}
\def\ea{\end{array}}
\def\bi{\begin{itemize}}
\def\ei{\end{itemize}}
\def\be{\begin{enumerate}}
\def\ee{\end{enumerate}}
\def\bc{\begin{center}}
\def\ec{\end{center}}
\def\bt{\begin{table}}
\def\et{\end{table}}
\def\btb{\begin{tabular}}
\def\etb{\end{tabular}}

\def\lsim{\raise0.3ex\hbox{$\;<$\kern-0.75em\raise-1.1ex\hbox{$\sim\;$}}}
\def\gsim{\raise0.3ex\hbox{$\;>$\kern-0.75em\raise-1.1ex\hbox{$\sim\;$}}}

\begin{document}
\hspace{11 cm}
\title{Anatomy of Higgs mass in Supersymmetric Inverse Seesaw Models}
\author{Eung Jin Chun}
\email{ejchun@kias.re.kr }
\affiliation{Korea Institute for Advanced Study, Seoul 130-722, Korea}
\author{V. Suryanarayana Mummidi}
\email{soori9@cts.iisc.ernet.in}
\author{ Sudhir K  Vempati}
\email{vempati@cts.iisc.ernet.in}
\affiliation{Centre for High Energy Physics, Indian Institute of Science,
Bangalore 560012}

\begin{abstract}
We compute  the one loop corrections to the CP even Higgs mass matrix  in the supersymmetric inverse seesaw model
to single out the different cases where the radiative corrections from the neutrino sector could become important. 
It is found that there could be a significant enhancement in the Higgs mass even for Dirac neutrino masses of 
$\mathcal{O}$(30) GeV if the left-handed  sneutrino soft mass  is comparable or larger than the right-handed neutrino mass. 
In the case where  right-handed neutrino masses are significantly larger than the supersymmety breaking scale, 
the corrections can utmost account to an upward shift of  3 GeV.  For very heavy multi TeV sneutrinos, the corrections replicate 
the stop corrections at 1-loop. We further show that general gauge mediation with inverse seesaw model naturally 
accommodates a 125 GeV Higgs with TeV scale stops.
\end{abstract}
\vskip .5 true cm

\pacs{73.21.Hb, 73.21.La, 73.50.Bk} 
\maketitle
\section{Introduction}
The discovery of the Higgs boson \cite{Aad:2012tfa, Chatrchyan:2012tx, Chatrchyan:2012ufa} of the standard model has put severe constraints on its supersymmetric extensions. In particular, for light stops masses (below 1 TeV),  maximal stop  mixing 
 is required to generate a  Higgs mass $\sim$125 GeV \cite{Hall:2011aa,Draper:2011aa,Christensen:2012ei,Aparicio:2012iw,Baer:2012uya,Arbey:2012dq,Cao:2012fz}. 
 Many supersymmetry breaking models 
 have already been strongly constrained by this requirement \cite{Heinemeyer:2011aa,AlbornozVasquez:2011aa,Arbey:2011aa,Arbey:2011ab,Carena:2011aa,Baer:2011ab,Kadastik:2011aa,Buchmueller:2011ab,Ellis:2012aa}.

On the other hand, neutrino masses constitute one of the strongest signatures of physics beyond standard model. 
It is imperative that any supersymmetric extension of the standard model should also contain an explanation for 
non-zero neutrino masses. Among many ideas to generate tiny neutrino massse,  the 
inverse seesaw model \cite{Mohapatra:1986bd} is interesting as it is applicable at the weak scale 
with neutrino Yukawa couplinig of order one, and thus testable at colliders like LHC.

In the present work, we revisit the consequences of the inverse seesaw model for the lightest CP even Higgs boson mass \cite{Gogoladze:2012jp,Guo:2013sna}. 
We find parameter regions in which  the one-loop corrections to the light Higgs mass can be very significant, 
leading to an increase of ${\cal O}(10)$ GeV, for the neutrino Yukawa coupling larger than about 0.2.  
This is in the line of observations of Refs. \cite{Babu:2008ge, Martin:2009bg} which explored the role 
of extra vector like matter at TeV scale  in increasing  the light Higgs mass. 
We then apply these corrections to phenomenological minimal supersymmetric standard models (PMSSM) 
and general gauge mediaed supersymmetry breaking models (GMSB) 
where the Higgs mass can become 125 GeV 
for supersymmetry breaking scale around TeV.

The paper is organized as follows. In the next section, we present one loop corrections to the Higgs mass and study the various parameter regimes. In the section 3, we work out two numerical examples in (1) PMSSM (2) General gauge mediated supersymmetric  inverse seesaw model. We conclude in the section 4.  Appendix A contains the
main  formulae, whereas appendices B and C contain RGE equations and some ancillary formulae.  

\section{One loop corrections to the Higgs mass in MSSM}

The inverse seesaw model  is characterized by a small lepton number violating mass, unlike the Type-I seesaw,  the right-handed neutrinos can be as light as  TeV or even below, with their Yukawa couplings of order one. 
This is achieved by having an additional singlet field, which we denote by S. The superpotential for this model is given as
\begin{equation}
W_{\text{SISM}} = W_{\text{MSSM}} + Y_N L H_u N^c + M_R N^c S + \mu_S SS 
\end{equation}
where $W_{MSSM}$ stands for the standard MSSM superpotential,
\begin{eqnarray}
W_{MSSM}= Y_U Q U^c H_u + Y_D Q D^c H_d  + Y_E L E^c H_d + \mu H_u H_d 
\end{eqnarray}
and $N^c$ and S are singlet fields carrying lepton number +1,-1 respectively.

We consider one right-handed neutrino and one singlet S field in the discussion. It can be easily generalized to the case of two/three generations. The mechanism of how neutrino gets mass is well documented in the literature. We revisit it here briefly. In the basis, $ \{{\nu}_L, {N^c},{S} \}$, the mass matrix, $M_{\nu}$, for the neutral leptons is given by
\begin{equation}
 M_{{\nu}} = \left( \begin{array}{ccc} 
0 & m_D & 0 \\
 m_D&0&  M_R\\ 
 0&M_R& \mu_S \end{array} \right)
\end{equation}
Where $ m_D=Y_N\, \langle H_u\rangle $. The eigenvalues are given as 
\begin{eqnarray}
 m_{\nu_1} &\approx&{m_D^2\, \mu_S\over M_R^2}\nn
 m_{\nu_2} &\approx&-\left({m_D^2\over 2\, M_R}+M_R\right)\nn
 m_{\nu_3} &\approx&\left({m_D^2\over 2\, M_R}+M_R\right)\nonumber
\end{eqnarray} 
 $ m_{\nu_1}$ is the lightest neutrino eigenvalue proportional to the lepton number violating parameter $\mu_S$, the other two eigenvalues are almost degenerate $\sim M_R$. 
 
 Since the inverse seesaw model is typically a low scale model, unlike the traditional seesaw mechanisms,  one
 wonders  if they can give large enough contribution to the light CP-even Higgs boson mass. This is more important 
 to explore in the regions where $m_D$ can be relatively large $\gtrsim 10 ~\text{GeV}$.  It should be noted that the 
 range in $m_D$ $\simeq(0.2-0.3) v$ for   has been explored by collider searches \cite{Bandyopadhyay:2012px,Banerjee:2013fga}.  There  are constraints, however,  
on the size of $m_D$ for a given value of $M_R$  from electroweak precision tests \cite{delAguila:2008pw}: 
  \begin{equation}
  \label{ewconstr}
 m_D \lesssim  0.05 ~M_R
\end{equation}
This constraint is strictly for the electron and muon generations. For the third generation, it is slightly weaker, at the level of 0.07.  
This requires $M_R\simeq 3 $ TeV for $m_D$ close to  the top quark mass.  To compute  the 
corrections to the light Higgs mass from the neutrino sector, we use the one loop effective potential methods of Coleman-Weinberg \cite{Coleman:1973jx}. The methods have been used to derive the well-known  one-loop corrections from the top-stop sector \cite{Ellis:1991zd, Drees:1991mx}
and we extend them to the neutrino sector in the inverse seesaw model. 

 The scalar potential in this model  consists of
\begin{equation}
V_S = V_F + V_D + V_{\text{soft}}
\end{equation}
where 
\begin{eqnarray}
V_F &=& | Y_e E H_d + Y_N H_u N|^2 + | Y_u Q u^c + \mu H_d + Y_N L N^c|^2 + |Y_N L H_u + M_R S|^2 + | M_R N^c + \mu_S S|^2 + \ldots \nonumber \\ 
V_D &=&\frac{1}{8}(g^2+g'^2)\,(|H_u|^2-|H_d|^2) \nonumber \\
V_{\text{soft}} &=& A_N L H_u N^c +B_M N S + B_{\mu_S} S S + H. c + \ldots  \,.
 \end{eqnarray} In the basis, $ \{\tilde{\nu}_L, \tilde{N^c}, \tilde{S} \} $, 
the mass matrix, $\mathcal{M}_{\tilde{\nu}}^2$, for the sneutrinos is given by
\begin{equation}
 \mathcal{M}_{\tilde{\nu}}^2 = \left( \begin{array}{ccc} 
 m_{L}^2 + D_L + m_D^2  & m_D\, ( A_N - \mu \cot \beta)  & M_R\, m_D \\
 *&  m_D^2 + m_{N}^2 + M_R^2 & B_M + M_R\, \mu_S \\ 
 *&*& M_R^2 + \mu_S^2+m_{\tilde{S}}^2 \end{array} \right) .
\end{equation}
In the above matrix, elements with $*$ correspond to symmetric entries of the mass matrix.  The eigenvalues of the
above mass matrix can be easily derived  in the limit $\mu_S \ll  m_D \ll M_R$, as required by the inverse seesaw
mechanism and the electroweak precision tests. In the leading order of $m_D M_R/d_2, m_D X_N/d_1 \ll 1$, 
they are given as 
\begin{eqnarray}
 m_{\tilde \nu_1}^2&\approx&m_L^2 + m_D^2 \left(1 + {M_R^2\over d_2 }+
     {X_N^2\over d_1}\right)\nn
 m_{\tilde \nu_2}^2&\approx& m_N^2 + M_R^2+m_D^2 \left(1-
       {X_N^2\over d_1}\right)\nn
 m_{\tilde \nu_3}^2&\approx&m_S^2 + M_R^2-
        {M_R^2\, m_D^2\over d_2}\nonumber
\end{eqnarray}
where
\begin{eqnarray}
d_1=m_L^2-m_N^2-M_R^2\nn
d_2=m_L^2-m_S^2-M_R^2 \,.\nonumber
\end{eqnarray}

One-loop corrections for the Higgs mass matrix will be derived from the one-loop effective scalar potential 
given by the standard form:
\begin{equation}
V_{1-loop} (q^2) = {1 \over 64 \pi^2} \mathcal{S}Tr \mathcal{M}^4 (h)~~ \text{Log} 
\left( {\mathcal{M}^2 (h) \over q^2} - {3 \over 2} \right) .
\end{equation} 
 In the basis $\Phi^T=(Re\{H_d^0\},Re\{H_u^0\})$,  the corrections to the CP even Higgs mass are given as 
 $$\mathcal{M}^2=\mathcal{M}^2_0+\Delta \mathcal{M}^2_t+\Delta \mathcal{M}^2_\nu$$ where
 $\mathcal{M}^2_0$ stands for the tree level mass matrix, $ \Delta \mathcal{M}^2_t$ and $\Delta \mathcal{M}^2_\nu$ are contributions
 from the top/stop sector and the neutrino/sneutrino sectors respectively.  The full mass matrix has the form:
 \begin{eqnarray}
\mathcal{M}_{11}^2&=& M_Z^2\,\cos\beta^2 + m_A^2 \sin\beta^2 +\Delta\mathcal{M}_{t_{11}}^2+ \Delta \mathcal{M_\nu}_{11}^2\nn
\mathcal{M}_{12}^2&=& -(M_z^2 + m_A^2) \cos\beta \sin\beta +\Delta\mathcal{M}_{t_{12}}^2 
 +\Delta \mathcal{M_\nu}_{12}^2\nn
\mathcal{M}_{22}^2&=&M_Z^2\, \sin\beta^2 + m_A^2 \cos\beta^2 +\Delta\mathcal{M}_{t_{22}}^2 
  +\Delta \mathcal{M_\nu}_{22}^2\nonumber
\end{eqnarray}
 where for the sake of completeness, we present the well known top/stop  contributions \cite{Ellis:1991zd, Drees:1991mx} :
 \begin{eqnarray}
\Delta\mathcal{M}_{t_{11}}^2&=& {3 g_2^2\,m_t^4\over 16 \pi^2 M_W^2 \sin\beta^2} \left(\mu \,X_t\over m_{\tilde t_1}^2- m_{\tilde t_2}^2\right)^2 g(m_{\tilde t_1}^2,m_{\tilde t_2}^2)\nn
\Delta\mathcal{M}_{t_{12}}^2&=& {3 g_2^2\,m_t^4\over 16 \pi^2 M_W^2 \sin\beta^2} \left(\mu \,X_t\over m_{\tilde t_1}^2- m_{\tilde t_2}^2\right) \log\left({m_{\tilde t_1}^2\over m_{\tilde t_2}^2}\right)- 
  {A_t\over \mu} \Delta\mathcal{M}_{t_{11}}^2\nn
\Delta\mathcal{M}_{t_{22}}^2&=& {3 g_2^2\,m_t^4\over 16 \pi^2 M_W^2 \sin\beta^2} \left(2\,\log{Q^2\over m_t^2 }+ 
        {2 A_t X_t\over m_{\tilde t_1}^2- m_{\tilde t_2}^2 }\log\left({m_{\tilde t_1}^2\over m_{\tilde t_2}^2}\right)\right) +\left ({A_t\over \mu}\right)^2 \Delta\mathcal{M}_{t_{11}}^2 
       \end{eqnarray}

 In the following we write down the contribution from the neutrino sector in a compact notation as follows:       
 \begin{eqnarray}        
\Delta \mathcal{M_\nu}_{11}^2&=& 2\,k\left(\tilde L_1\, \tilde B_{11}^2+\tilde L_2\,\tilde B_{21}^2+\tilde L_3\,\tilde B_{31}^2+ m_{\tilde \nu_1}^2 \tilde A_{111}(\tilde L_1-1)+     m_{\tilde \nu_2}^2 \tilde A_{211}(\tilde L_2-1)+m_{\tilde \nu_3}^2 \tilde A_{311}(\tilde L_3-1)\right)\nn
\Delta \mathcal{M_\nu}_{22}^2&=& 2\,k\left(\tilde L_1\,\tilde B_{12}^2+\tilde L_2\,\tilde B_{22}^2+\tilde L_3\,  \tilde B_{32}^2+  m_{\tilde \nu_1}^2 \tilde A_{122}(\tilde L_1-1) +
 m_{\tilde \nu_2}^2 \tilde A_{222}(\tilde L_2-1)+m_{\tilde \nu_3}^2 \tilde A_{322}(\tilde L_3-1)\right)\nn
&-&4\,k\,\left(L_{2}\, B_{22}^2+L_{3}\, B_{32}^2+m_{ \nu_2}^2  A_{222}( L_2-1)+m_{ \nu_3}^2  A_{322}( L_3-1)\right)\nn
\Delta \mathcal{M_\nu}_{12}^2&=& 2\,k\left(\tilde L_1\,\tilde B_{12} \tilde B_{11}+\tilde L_2\,\tilde B_{22} \tilde B_{21}+\tilde L_3\,\tilde B_{32} \tilde B_{31}+  m_{\tilde \nu_1}^2 \tilde A_{112}(\tilde L_1-1) +
m_{\tilde \nu_2}^2 \tilde A_{212}(\tilde L_2-1)  \right. \nn
&+& \left. m_{\tilde \nu_3}^2 \tilde A_{312}(\tilde L_3-1)\right)
\label{higgscor}
\end{eqnarray}       
In the above, 
\begin{eqnarray}
g(A,B) &=& 2 -{ (A+B )\over(A- B)} \log {A\over B};~~~~
k=\frac{2}{32\pi^2}\nn
\tilde L_i&=& Log\left( \frac{m_{\tilde\nu_i}^2}{Q^2}\right);~~~~~~~~~~
L_i=Log\left(\frac{m_{\nu_i}^2}{Q^2}\right)\nn
X_t&=& A_t-\mu \cot\beta\nonumber
\end{eqnarray}
and  $B_{ij}$ and $A_{ijk}$'s are given in the Appendix.  While the above formulae are written for
a single generation of right handed and singlet neutrinos,  they can be 
easily generalized to three generations of right-handed and singlet neutrinos.
 The neutrino contributions to the light Higgs mass, though similar to those from the top/stop
 sector, have a couple of distinct features: (a) there is no colour factor associated with the
 neutrino contributions, so they typically lower than the top/stop contributions by a factor three,
 (b) The fermionic contributions, from the right-handed neutrinos can be significant, reducing
 the total contribution to the Higgs mass.  This is highly dependent on the hierarchies between the
 relevant parameters: the soft masses and the right handed neutrino masses. 
   To understand the overall relevance
 of these contributions, we will consider a few interesting cases below. Note that  
 in our numerical analysis, we restrict $m_D M_R/d_2, m_D X_N/d_1$ to be less than 0.1.

\subsubsection{case-1 : $M_R \approx m_L$}
\begin{figure}[h]
\includegraphics[width=0.45\textwidth,angle=0]{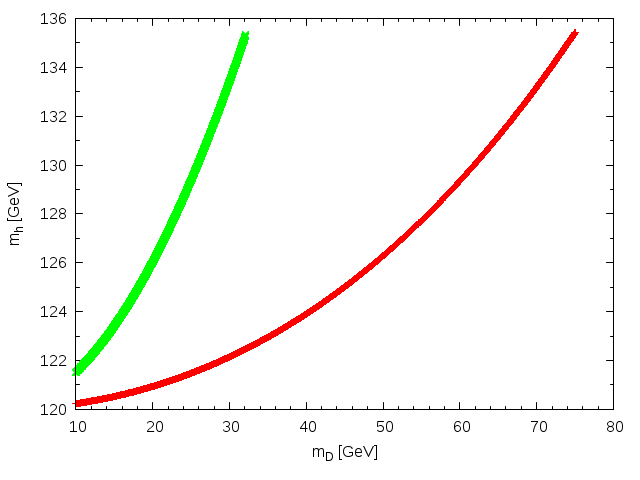}
\caption{The lightest CP even Higgs boson mass as a function of $m_D$ for $A_N=0$ (red) and $A_N=-1000$ GeV (green). In this plot other parameter are fixed as $M_R= m_L=1500$ GeV, $m_N=1000$ GeV and $m_S=800$ GeV. The stop mass parameters are fixed such that $m_h = 120$ GeV without the neutrino corrections.}
\label{Higgsvsall}
\end{figure}
In this case \footnote{see Appendix-C, for approximate formulae for sneutrino eigenvalues},  we choose the 
right-handed neutrino mass scale close to the (left-handed) slepton masses.  To satisfy the electroweak
precision tests, $m_D$ should be typically smaller than $M_R$ by a factor 20.  Neutrino masses can be adjusted by choosing a 
sufficiently small $\mu_S$. In Fig.~\ref{Higgsvsall} we plot the light Higgs mass as a function of $m_D$ for two values of $A_N$ = 0 (Red) and $A_N =  - 1000$ GeV (Green). The stop contributions are chosen such  that the lightest CP even Higgs mass, $m_h = 120\, \text{GeV }$
and rest of the contribution comes solely from the neutrino sector. As it can be seen from the left panel, the Higgs mass has a significant increase from
120 GeV and the increase is possible even for $m_D$ values as small as 20 GeV \footnote{$m_D$ values quoted here are at the $M_{SUSY}$ scale
which has been fixed at 1 TeV.} as long as slepton mass $m_L$ is relatively heavy $\gtrsim$ 1TeV in the same range of $M_R$. The rest of the
slepton masses appearing in the 1-loop formula are chosen to be $m_N = 1000$ GeV and $m_S  = 800$ GeV.  As expected increasing $A_N$ increases the higgs mass, but the effect varies with increasing $m_D$.   
It should be noted that perturbative constraints exist on the neutrino Yukawa couplings  $y_N  = m_D /v_u $. Requiring $y_N$ to be
perturbative all the way up to the GUT scale, puts a constraint on $y_N \leq 0.75$ \cite{Guo:2013sna}.  
While we have considered $M_R \approx m_L$ in the present example, this is not strictly necessary.  
For example, instead of $m_L$ in the above case, one can have a similar enhancement
in the case $M_R \approx m_N$, while $m_L$ being significantly lighter. $M_R \approx m_S$ does not 
lead to significant enhancement because that $S$ does not  couple to the Higgs field, $H_u$.  
\begin{figure}[h]
\includegraphics[width=0.45\textwidth,angle=0]{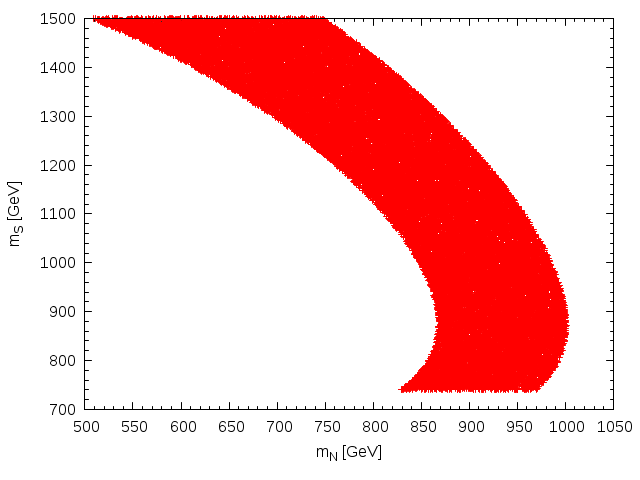}
\caption{ The red band corresponds to Higgs mass of range [124-126] GeV in the plane
of $m_S$ and $m_N$. The rest of the parameters are chosen as $m_L  =  M_R = 1500$ GeV, $m_D=75$ GeV and $A_N=0$. The stop mass parameters are fixed such that $m_h = 120$ GeV without the neutrino corrections.}
\label{mnms}
\end{figure}
In Fig.~\ref{mnms}, $m_L$, $m_D$ and $A_N$ are fixed to be 1500 GeV, 75 GeV and 0 respectively. And the parameter space in $m_N$ and $m_S$ plane is plotted with a restriction that Higgs mass should be in the range [124-126] GeV. Evidently there is wide range of parameter space available which can give Higgs mass of 125 GeV.

%\begin{figure}[h]
%\includegraphics[width=0.30\textwidth,angle=0]{mh-Ml.png}
%\caption{ $m_h$ vs $m_L$ is plotted for two different values for $A_N$ while fixing the other parameters.  Legend for the plots are as follows Red:\,$A_N$=0,$m_N$=500 GeV,$m_S$=400 GeV
%Green:\,$A_N$=-1000 GeV,$m_N$=500 GeV,$m_S$=400 GeV}
%\label{ynmr}
%\end{figure}
%

\subsubsection{case-2: $M_R \gg m_L$}

\begin{figure}[h]
\includegraphics[width=0.45\textwidth,angle=0]{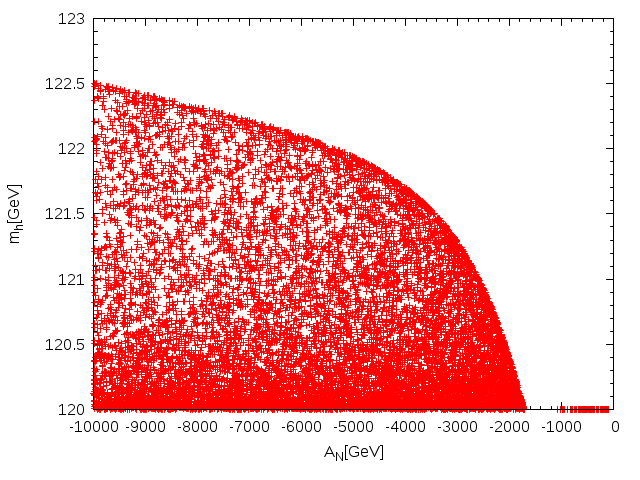}
\includegraphics[width=0.45\textwidth,angle=0]{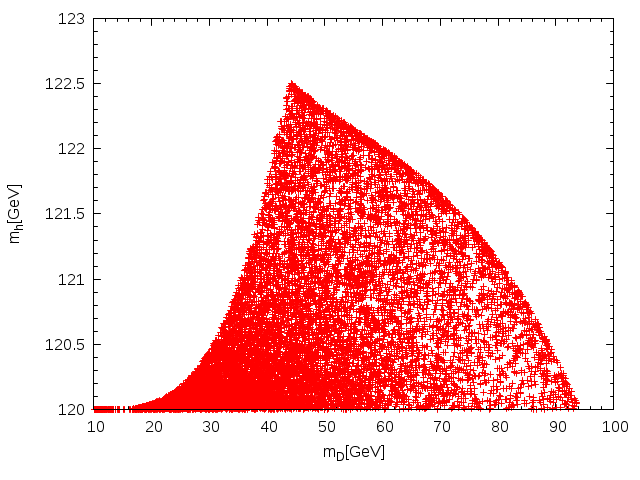}
\caption{The lightest CP even neutral Higgs boson is plotted for\,$M_R=5$ TeV with respect to $A_N$ (left panel) and  $m_D$ (right panel).
 The rest of the parameters are chosen to be   $m_L=500$ GeV, $m_N=300$ GeV, and $m_S=200$ GeV. 
The stop parameters are taken such that
  $m_h = 120$ GeV. }
\label{mrlargecase2}
\end{figure}

We now consider the case where $M_R$ is the largest mass scale in the theory. This limit has been earlier considered in 
Ref.\cite{Guo:2013sna}. In this case, the enhancement in the light Higgs mass is much smaller and restricted to a few GeV.  This is because the neutrino 1-loop correction, which is negative, significantly suppresses the total 
contribution from the neutrino sector.  This is illustrated in Fig.~\ref{mrlargecase2} where we have plotted $m_h$ as a
function of $A_N$ (left panel) and $m_D$ (right panel). The slepton masses are fixed as $m_L =  500 $ GeV, $ m_N = 300$ GeV
and $m_S = 200$ GeV.  The right-handed neutrino mass is taken to be 5 TeV.  As can be seen from the plots, the 
enhancement is not significant in this case. This holds  even with the variation in $A_N$ (left panel) or $m_D$ (right panel). 
The maximum enhancement achieved here is about two and half GeV. 

\subsubsection{case-3: $  m_L = m_N = m_S \gg M_R $}

\begin{figure}[h]
\includegraphics[width=0.45\textwidth,angle=0]{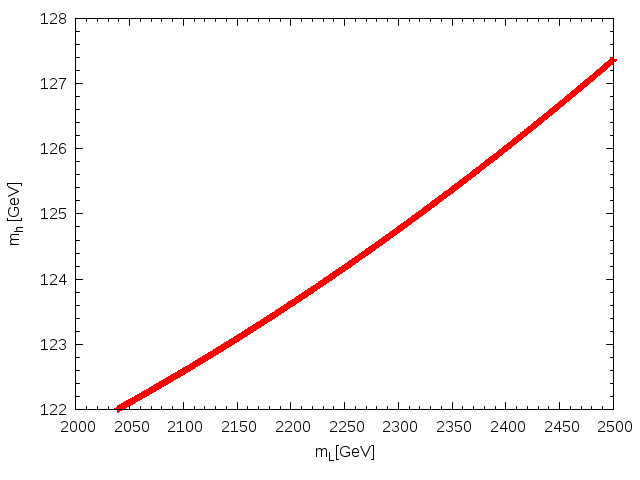}
\caption{$m_h$ is plotted versus $m_L$  keeping $M_{R}=500$ GeV and $A_N=0$ . The stop parameters are taken such that
  $m_h = 120$ GeV.}
\label{mrsmallcase3}
\end{figure}

This case replicates the stop corrections. All the sneutrino eigenvalues are much larger than the neutrino ones and thus dominating over
the negative contributions.  However, it turns out that the required sneutrino mass scale is in  TeV range ( around 2 TeV range for a 
500 GeV $M_R$). This range is suited
for semi-split and split scenarios. This is depicted in the Fig. \ref{mrsmallcase3}. 

\subsubsection{case-4: $ M_R = m_{SUSY} $}

\begin{figure}[h]
\includegraphics[width=0.45\textwidth,angle=0]{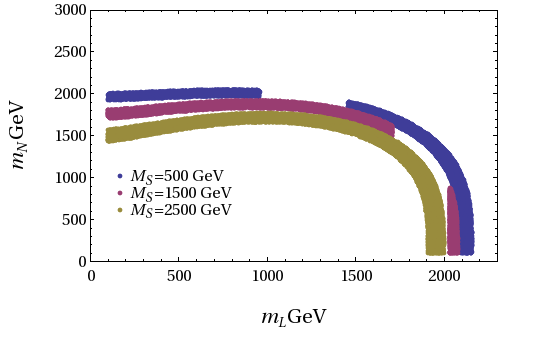}
\includegraphics[width=0.45\textwidth,angle=0]{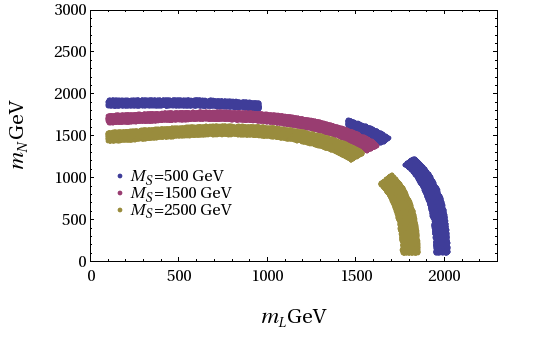}
\caption{$m_L$ is plotted against $m_N$ with different values of $m_S$, for a Higgs mass of 124-126 GeV. 
The rest of the parameters are chosen to be   $A_N=0$ (-1 TeV) and $m_D=50$ GeV for the left (right) panel. The stop parameters are taken such that
  $m_h = 120$ GeV}
\label{mrmsusy}
\end{figure}

We now consider the case where $ M_R ~= M_{SUSY} $, where $M_{SUSY} = \sqrt{m_{\tilde{t}_1} m_{\tilde{t}_2}}$. 
 Fixing the stop parameters such that $m_h = 120 $ GeV fixes,  $M_R \approx M_{SUSY} \approx 1 $TeV.  $m_N, m_L$ and $m_S$ are considered
 as free parameters. The parameter space $m_L$ and $m_N$ which accommodates, Higgs mass in the range 124--126 GeV for different values of $m_S$ 
 with $A_N=0$ and $A_N = -1 $ TeV is shown in Fig.~\ref{mrmsusy}.\footnote {Points which contribute to poles in sneutrino eigenvalues are removed and are responsible for discontinuity in the plots} 

\section{Applications to PMSSM and GMSB} 

In the present section, we present two numerical examples as an application to the above calculation.
\subsection{PMSSM and Inverse Seesaw}

\begin{figure}[h]
\includegraphics[width=0.45\textwidth,angle=0]{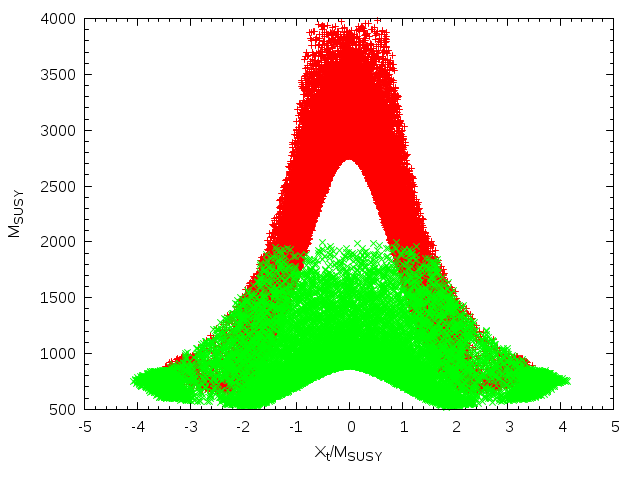}
\includegraphics[width=0.45\textwidth,angle=0]{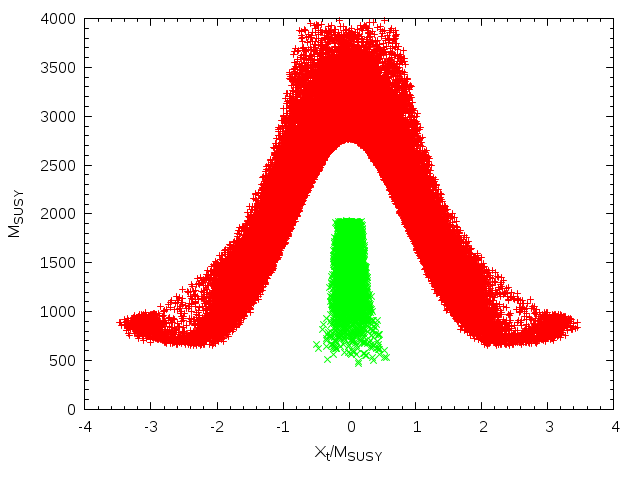}
\caption{For a fixed value of $M_R=1000\, (2500)$ GeV  in the left (right) panel, $M_{SUSY}$ is plotted against $X_t/ M_{SUSY}$. Green (red) points are with (without) neutrino contribution.}
\label{pmssm}
\end{figure}
%In the first case, we consider PMSSM like model with 9 parameters. The following choices are made for the ranges of the parameters.\\
%~~~~$m_Q^2,m_U^2,m_L^2,m_N^2$,$m_S^2$ $\rightarrow$\,[100\,,\,3000]\,Gev;~~~ $A_t,\mu$$\rightarrow$\,[-1500\,,\,1500]\,Gev\\
%~~~~~$Y_N$\,\,$\rightarrow$\,[0.1\,,\,1];~~~~~$M_R$\,$\rightarrow$\,$[10^2\,,\,10^4]$\,Gev
%In Fig\ref{xthiggs}, we show the impact of adding the contribution from the neutrinos. In Fig2a, we show the famous bell curve of the Higgs mas in the $M_{SUSY}$,$X_t\over M_{SUSY}$ plane. The red points have no'SISM"  contribution, where as the green points are with "SISM" contributions. Both the red points and the green points have Higgs mass in the range [123,128]\, Gev. As it can be seen from the plots, a 125 Gev Higgs from light stops $\simeq$ O(1Tev) is possible even with $X_t$$\simeq$0. We have fixed $M_R=500Gev$, for Fig2a.\\
%More explicitly, we see the impact of the neutrinos when $X_t\simeq0$ and $M_{SUSY}\simeq (1-1.5)$Tev. In Fig2b, we plot these points. As can be seen a 125 Gev Higgs can be easily organised with a reasonbly large $Y_N\simeq$ 0.2-0.8
%
The phenomenological MSSM is  low energy parameterisation of the supersymmetry breaking soft terms in terms of 19-22 parameters (See for example, \cite{Berger:2008cq}). 
To study the inverse seesaw model in the PMSSM setting the following additional parameters $m_N, m_S, A_N$ are to be included.
Together with the existing parameters, the nine parameters which completely fix the low energy neutral  CP even Higgs mass matrix and
their ranges are given as: $m_Q, m_U, m_L, m_N, m_S \in[100, 3000]$ GeV;  
$A_t, \mu \in [-1500, 1500]$GeV;  $Y_N \in [0.1, 1]$.

In Fig.~\ref{pmssm}, we have plotted the regions of Higgs mass within the range   [123,128]\, GeV in the plane of $M_{SUSY}$ and $X_t/M_{\text{SUSY}}$. 
In the left (right) panel right-handed neutrino mass is chosen to be 1 (2) TeV.
 The $\tan\beta$ is fixed to be 10. The red points are the ones without neutrino/sneutrino contributions 
and the green points are the ones where neutrino/sneutrino contributions are added.  
As we can clearly see from the figure, 
even if $m_{SUSY}$ is below 1 TeV, there is enough contribution from the neutrino sector to a 125 GeV Higgs mass 

As right-handed neutrino mass $M_R$ increases, we get closer to the case-2 discussed in the previous section where larger right-handed neutrino masses make the contribution of right-handed neutrinos to Higgs mass negative, and thus reducing the Higgs mass. This effect is seen in the right panel of Fig.~\ref{pmssm}.

\subsection{GMSB and Inverse Seesaw}

Minimal gauge mediation models have been strongly constrained by the recent discovery of the Higgs mass of 125 GeV \cite{Draper:2011aa}.
This is because the stop mixing parameter $X_t$ is predicted to be very small in these models.  The $X_t$ can be made large through 
renormalisation group corrections, but this would require gluino masses to be greater than 8 TeV. Thus, the only way these models can accommodate
a light CP even neutral higgs boson with a mass around 125 GeV is by increasing the masses of the stops beyond 4 TeV. This range for the stop masses
is far beyond the LHC reach. 
One can then consider modification by including either messenger-matter interactions,  new fields  or 
 new interactions to achieve the Higgs mass within the required ball park.  
The feature of small $X_t$ also persists in general gauge mediation which is an umbrella of all possible gauge mediations both in the perturbative
and non-perturbative regime. A recent analysis of the Higgs mass in general gauge mediation is presented in Ref. \cite{Grajek:2013ola}. 

Incorporating the inverse seesaw model in general gauge mediation could generate the Higgs mass in the right ball park, due to the additional corrections induced by the neutrinos. We study this possibility in the present subsection.
The set up of general gauge mediation we consider is specified by the following boundary conditions at the messenger scale:
\begin{eqnarray} 
M_i (X)&\approx&  \frac{\Lambda}{16\pi^2} \sum_i \left(g_i^2(X) \right) \nonumber  \\
m_{\tilde Q}^2 &\approx& \frac{2\Lambda^2}{(16\pi^2)^2}\sum_i \left(g_i^4 (X) ~C_i (Q) \right)  \nonumber \\
m_{\tilde U}^2 &\approx& \frac{2\Lambda^2}{(16\pi^2)^2}\sum_i \left(g_i^4 (X) ~C_i (U) \right) \nn
m_{\tilde L}^2 &\approx& \frac{2\Lambda_L^2}{(16\pi^2)^2}\sum_i \left(g_i^4 (X) ~C_i (L) \right) \nn
m_{\tilde e}^2 &\approx& \frac{2\Lambda_L^2}{(16\pi^2)^2}\sum_i \left(g_i^4 (X) ~C_i (L) \right) \nn
m_{\tilde S}^2 &=&0\nn
m_{\tilde N}^2 &=&0\nonumber
\end{eqnarray}
Where $C_i(f)$ is the quadratic casimir of the field f and $g_i(X)$ is the gauge coupling constant at the messenger scale $M_X$. And 'i' runs over all the gauge groups in the standard model. Except for sleptons all other parameters, scalar squared masses and gaugino masses are set by $\Lambda$ at the messenger scale where as slepton masses are 
set by $\Lambda_L$.
Typically, the soft masses of the singlets $m_S $ and $m_N$ are zero at the mediation scale. 
At the weak scale $m_N$ does get generated by RGE corrections whereas $m_S$ remains zero. 
Using the RGE given in the appendix, the leading log estimate of $m_N$ at the 
weak scale is given by 
\begin{equation}
m_N^2 (M_\text{SUSY}) \approx  - {1 \over 16 \pi^2} ( m_{H_u}^2 + m_L^2 + m_N^2)  Y_N^2  Log ({M_{\text{mess}} \over M_{\text{SUSY}} })
\end{equation}
This generates a large enough positive contribution to $m_N^2$ at the weak scale as $m_{H_u}^2$ is negative at the weak scale from
the requirement of electroweak symmetry breaking.  The question then remains whether with the above boundary conditions it  is
possible to reproduce either of the conditions $m_N \gtrsim M_R $ or $m_L \gtrsim M_R$ to enhance the Higgs mass 
significantly.

Assuming as before only one right-handed neutrino and one singlet, we find that it is indeed possible to generate a Higgs mass
of 125 GeV.  We have to choose an appropriate boundary condition for the third generation sleptons such that it is close to the
$M_R$ mass. In the table \ref{ex1} we present two example points which have this characteristic. It is clear that the two example points given in the table corresponds to the case $m_L$$\sim$ $M_R$ discussed in the section-2. 
\begin{table}
\begin{tabular}{|c| c| c|c | c|c|c|c|c|c|c|}
\toprule
$m_h[GeV]$& $y_N$ & $M_R[GeV]$ & $ m_{\tilde t_1}$[GeV]& $m_{\tilde t_2}$ [GeV]& $\Lambda$[GeV]&$\Lambda_L$[GeV]&$m_L$[GeV]&$m_N$[GeV]&$m_S$[GeV]\\ \hline
127.21& 0.62 & 1787 & 943 & 1078& $10^5$& 5.2 $10^5$& 1791 &808 & 0 \\\hline
124 & 0.82& 2056& 940 &1072 &$10^5$& 6 $10^5$ & 2046 & 846 &0 \\ \hline
\end{tabular}
\caption{Parameter space in general gauge mediated supersymmetric inverse seesaw model}
\label{ex1}
\end{table}

\section{Summary}

Inverse seesaw model has many interesting features and serves as an important alternative to the regular seesaw model. 
Supersymmetric versions of this model have been studied earlier in the literature. In the present work, we have discussed
the detailed anatomy of the one loop corrections to the neutral CP even Higgs boson masses. We show that the corrections
can be significant in cases where the soft mass of either the singlet or the doublet sneutrino is comparable or greater than
the right-handed neutrino mass (for reasonable values of Dirac coupling).  An enhancement of 6-12 GeV or even more 
can be easily achieved. This removes the requirement of a large stop mixing parameter $X_t$ (for stop masses less than
a TeV) in models where low scale inverse seesaw mechanism is implemented. 

An interesting application of this model lies in general gauge mediation where we have shown that implementing inverse
seesaw model can enhance the light Higgs mass to the 125 GeV for stops less than a TeV, without resorting to any
mechanism to enhance the stop mixing parameter $X_t$. 

\bigskip

\noindent 
\textbf{Acknowledgments:}\\
 SKV is supported by DST Ramanujan Grant SR/S2/2008/RJN-25 of Govt of India. VSM is supported by CSIR fellowship 09/079(2377)/2010-EMR-1. EJC was supported by SRC program of NRF Grant No.\ 2009-0083526
funded by the Korea government (MSIP) through Korea Neutrino Research Center.

\appendix 

\section{Appendix}
Here we have collected the expressions for $B_{ij}$ and $A_{ijk}$'s which are used in the calculation of one-loop corrected Higgs mass.
\begin{eqnarray}
B_{ij}&=&{\partial m_{\nu_i}^2 \over \partial H_j}\nn
A_{ijk}&=&{\partial B_{ij} \over \partial H_k}\nonumber
\end{eqnarray}

\begin{equation}
\tilde B_{12}=2\,y_N\, m_D\left(\frac{m_L^2-m_S^2}{d_2}+\frac{X_N A_N }{d_1}\right)\nonumber ;~~~\tilde B_{11}=-2\,y_N\, m_D\frac{\mu\, X_N}{d_1}\nonumber;~~~~
\tilde B_{22}=2\, y_N\, m_D-2\, y_N\, m_D\,\frac{X_N A_N }{d_1}\nn
\end{equation}
\begin{equation}
\tilde B_{21}={2\, y_N\, m_D\,\mu\,X_N\over d_1}\nonumber ;~~~
\tilde B_{32}=-{2\, y_N\, m_D\,M_R^2\over d_2}\nonumber;~~~
\tilde B_{31}=0
\end{equation}
\begin{eqnarray}
\tilde A_{111}&=&\frac{2 y_N^2 \mu^2}{d_1};~~~~~~
\tilde A_{112}=-{A_N\over \mu}\,\tilde A_{111};~~~~~
\tilde A_{122}=2 y_N^2 (1+{A_N\over d_1}+{M_R^2\over d_2})\nn
\tilde A_{211}&=&-\tilde A_{111}:~~~~~
\tilde A_{212}={A_N\over \mu}\,\tilde A_{111};~~~~~~~
\tilde A_{222}=2 y_N^2 (1-{A_N\over d_1})\nn
\tilde A_{311}&=&0:~~~~~~~~~~~~\tilde A_{312}=0:~~~~~~~~~~~~~~~~\tilde A_{322}=-2 y_N^2{ M_R^2\over d_2}\nn
B_{22}&=&2|H_u|y_N^2+\frac{|H_u|^3 y_N^4}{M_R^2};~~~~~~~~
B_{32}=2|H_u|y_N^2+\frac{|H_u|^3 y_N^4}{M_R^2}\nn
A_{222}&=&2 y_N^2+\frac{3|H_u|^2 y_N^4}{M_R^2};~~~~~~~~~~~~~~~~~~~~~
A_{322}=2 y_N^2+\frac{3|H_u|^2 y_N^4}{M_R^2}\nonumber
\end{eqnarray}
where we have suppressed the generation indices. 

\section{RGE equations in SISM }
In the last section of the appendix we present the renormalisation group equations for some of the superpotential
and soft terms relevant to the analysis of general gauge mediation. To derive the formulae we use the standard formulae available
in the literature\cite{Falck:1985aa,Martin:1993zk}.  The notation we use is $t=Log({\mu\over m_{SUSY}})$. 
\begin{eqnarray}
\frac{d y_i}{dt} &=& \frac{y_i}{16 \pi^2}{\gamma_i}^{(1)}\nn
\frac{d \mu}{dt} &=& \frac{\mu}{16 \pi^2}\left[3 y_t^2+3 y_b^2+y_N^2+y_{\tau}^2-3 g_2^2-{3 \over 5}g_1^2\right]\nn
\frac{d \mu_s}{dt} &=& 0\nn
\frac{d M_R}{dt} &=&\frac{M_R}{16 \pi^2} 2 y_N^2\nonumber
\end{eqnarray}
where
\begin{eqnarray}
\cr{\gamma_t}^{(1)}&=&\left[y_N ^2+6 y_t^2 + y_b^2-{16\over 3} g_3^2 - 3 g_2^2 - {13\over 15} g_1^2\right] \nn
\cr{\gamma_b}^{(1)}&=&\left[6 y_b^2 + y_t^2+{y_\tau}^2-{16\over 3} g_3^2 - 3 g_2^2 - {7\over 15} g_1^2 \right] \nn
\cr{\gamma_{\tau}}^{(1)}&=&\left[y_N ^2+3 y_b^2 +4{y_\tau}^2 -3 g_2^2- {9 \over 5} g_1^2\right] \nn
\cr{\gamma_{y_N}}^{(1)}&=&\left[4 y_N ^2+3 y_t^2+{y_\tau}^2-{3 \over 5}g_1^2 \right] \nonumber
\end{eqnarray}

\begin{eqnarray}
\frac{d m_{H_u}^2}{dt} &=&\frac{1}{16 \pi^2}\left[3 x_t +x_N- 6 g_2^2 M_2 -\frac {6}{5}g_1^2 M_1^2+ {3\over 5}g_1^2 \xi\right]\nn
\frac{d m_{H_d}^2}{dt} &=&\frac{1}{16 \pi^2}\left[3 x_b +x_{\tau}- 6 g_2^2 M_2 -\frac {6}{5}g_1^2 M_1^2- {3\over 5}g_1^2 \xi\right]\nn
\frac{d m_{N}^2}{dt} &=&\frac{1}{16 \pi^2}\left[ 2\, x_N\right]\nn
\frac{d m_{S}^2}{dt} &=&0
\end{eqnarray}
where
\begin{eqnarray}
x_t &=& 2 y_t^2 \left(m_{H_u}^2+m_{Q_3}^2+m_{U_3}^2\right)+2 A_t^2 \nn
x_b &=& 2 y_b^2 \left(m_{H_d}^2+m_{Q_3}^2+m_{d_3}^2\right)+2 A_b^2\nn
x_{\tau} &=& 2 y_{\tau}^2 \left(m_{H_d}^2+m_{L_3}^2+m_{e_3}^2\right)+2 A_{\tau}^2\nn
x_N &=& 2 y_N^2 \left(m_{H_u}^2+m_{L_3}^2+m_{N}^2\right)+2 A_N^2\nn
\end{eqnarray}
\section{Approximate sneutrino egenvalues for $M_R \sim m_L$}
\begin{eqnarray}
m_{\tilde \nu_1}^2&\approx&m_L^2 + m_D^2 \left(1-\frac{m_L^2}{m_S^2}-\frac{X_N^2}{m_N^2}\right)\nn
m_{\tilde \nu_2}^2&\approx& m_N^2 + m_L^2+m_D^2 \left(1+
\frac{X_N^2}{m_N^2}\right)\nn
m_{\tilde \nu_3}^2&\approx&m_S^2 +m_L^2+m_D^2 \,\frac{m_L^2}{m_S^2}\nonumber\\
\tilde B_{11}&=&2\,y_N\,m_D\,\frac{\mu\,X_N}{m_N^2}\nonumber ;~~~\tilde B_{21}=-\tilde B_{11}\nonumber\\
\tilde B_{12}&=& 2\,y_N\,m_D\,\left(1-\frac{m_L^2}{m_S^2}-\frac{X_N\,A_N}{m_N^2}\right)\nn
\tilde B_{22}&=& 2\,y_N\,m_D\,\left(1+\frac{X_N\,A_N}{m_N^2}\right)\nn
\tilde B_{32}&=&2\,y_N\,m_D\frac{m_L^2}{m_S^2}\nonumber
\end{eqnarray}
\begin{eqnarray}
\tilde A_{111}&=&-{2 \mu^2 y_N^2\over m_N^2 };~~~~~~
\tilde A_{112}=-{A_N\over \mu} \,\tilde A_{111}\nn
\tilde A_{122}&=&2 y_N^2 (1 - {A_N^2\over m_N^2} - {M_R^2\over m_S^2});~~~~~~~~~~~~~~~~~~~
\tilde A_{211}=-\tilde A_{111}\nn
\tilde A_{212}&=&{A_N\over \mu} \,\tilde A_{111};~~~~~~~~~~~~~~~~~~~~
\tilde A_{222}=2 y_N^2 (1 + {A_N^2\over m_N^2})\nn
\tilde A_{322}&=& 2 y_N^2 {M_R^2\over m_S^2}
\end{eqnarray}
Approximate formula for the bound on the mass of the lightest Higgs boson is given by
\begin{eqnarray}
m_h^2&\le& M_Z^2 \cos\beta^2+\Delta \mathcal{M_t}_{22}^2 \sin\beta^2+\Delta \mathcal{M_\nu}_{22}^2 \sin\beta^2
\end{eqnarray}
$\Delta \mathcal{M}_{11}^2$ and $\Delta \mathcal{M}_{12}^2$ contribution is very small, compared to $\Delta \mathcal{M}_{22}^2$, can be neglected. To estimate nuetrino contribution to Higgs mass we have considered $\Delta \mathcal{M_\nu}_{22}^2$ in the approximations $M_R\sim m_L$ and $A_N$=0, which is given by
\begin{eqnarray}
\Delta \mathcal{M_\nu}_{22}^2&=& 2\,k\left(\tilde L_1\,\tilde B_{12}^2+\tilde L_2\,\tilde B_{22}^2+\tilde L_3\,  \tilde B_{32}^2+ m_{\tilde \nu_2}^2 \tilde A_{222}(\tilde L_2-1)+m_{\tilde \nu_3}^2 \tilde A_{322}(\tilde L_3-1)\right)\nn
&&-4\,k\,\left(L_{2}\, B_{22}^2+L_{3}\, B_{32}^2+m_{ \nu_2}^2  A_{222}( L_2-1)+m_{ \nu_3}^2  A_{322}( L_3-1)\right)
\label{h22}
\end{eqnarray}
From Appendix-A, it is clear that A's and B's corresponding to right-handed neutrino part is are small compared to that of sneutrinos. Thus fermion contribution can be safely neglected (this is not true when $m_{SUSY}$$\ll$$M_R$) and \ref{h22} becomes 

\begin{eqnarray}
\Delta \mathcal{M_\nu}_{22}^2\over (2 k)&\approx& 4m_D^2 y_N^2\left[\tilde L_2+{M_R^4\over m_S^4}L_3+{(M_R^2-m_S^2)^2\over m_S^4} \tilde L_1\right]\nn
&&+ 2 m_{\tilde\nu_1}^2 y_N^2\left[({M_R^2 \over m_S^2}-1)\,(1-\tilde L_1)\right]+ 2 m_{\tilde\nu_2}^2 y_N^2\left[\tilde \tilde L_2-1\right]\nn
&&+ 2 m_{\tilde\nu_3}^2 y_N^2\left[{M_R^2 \over m_S^2} \,(\tilde L_3-1)\right]
\end{eqnarray}
Typically all the sneutrino masses are of O($10^6$) (while log factors are of O(1)) and can be taken to be equal.
\begin{eqnarray}
\Delta \mathcal{M_\nu}_{22}^2\over (2 k)&\approx& 4m_D^2 y_N^2\left[\tilde L_2+{M_R^4\over m_S^4} \tilde L_3+{(M_R^2-m_S^2)^2\over m_S^4}\tilde L_1\right]
+2 \,m_{\tilde\nu}^2\, y_N^2\left[{M_R^2\over m_S^2}\,{\tilde L_3 \over \tilde L_1}\right.\nn
&&\left.+(-2+\tilde L_1+\tilde L_2)\,\right]
\label{meq}
\end{eqnarray}
From eq (\ref{meq}), it is evident that Higgs mass receives large correction from sneutrino masses. As sneutrino masses implicitly depend on $m_D$, increase in $m_D$ increases the Higgs mass.
\bibliographystyle{ieeetr}
\bibliography{seesaw.bib}
\end{document}